\documentclass[twocolumn,showpacs,prb]{revtex4}

\usepackage{graphicx}

\begin{document}

\title{A Role of Initial Conditions in\\
Spin-Glass Aging Experiments}
\author{V.~S.~Zotev}
\author{G.~F.~Rodriguez}
\author{R.~Orbach}
\affiliation{Department of Physics, University of California,
Riverside, California 92521}
\author{E.~Vincent}
\author{J.~Hammann}
\affiliation{Service de Physique de l'Etat Condense, CEA Saclay,
91191 Gif-sur-Yvette Cedex, France}
\date{Submitted to PRB on February 22, 2002}

\begin{abstract}
Effect of initial conditions on aging properties of the spin-glass
state is studied for a single crystal Cu:Mn 1.5 at \%. It is shown
that memory of the initial state, created by the cooling process,
remains strong on all experimental time scales. Scaling properties
of two relaxation functions, the TRM and the IRM (with $t_{w1}=0$
and $t_{w2}=t_{w}$), are compared in detail. The TRM decay
exhibits the well-known subaging behavior, with $\mu<1$ and
$t_{w}^{e\!f\!f}>t_{w}$. The IRM relaxation demonstrates the
superaging behavior, with $\mu>1$ and $t_{w}^{e\!f\!f}<t_{w}$. It
is shown that an average over different initial conditions leads
to systematic improvement in $t/t_{w}$ scaling. An effective
barrier, describing influence of the initial state on spin-glass
relaxation, is found to be almost independent of temperature.
These results suggest that departures from full $t/t_{w}$ scaling,
observed in aging experiments, are largely due to cooling effects.
\end{abstract}


\maketitle

\section{\label{intro}Introduction}

Aging phenomena in spin glasses have been studied for almost
twenty years.\cite{lun83} Despite considerable progress, some
problems remain open. One of them is a problem of $t/t_{w}$
scaling of time dependent quantities.\cite{vin96} This issue is a
touchstone of our understanding of spin-glass dynamics. Only when
all departures from full $t/t_{w}$ scaling are accounted for, both
theoretically and experimentally, can one say that the aging
phenomena are well understood.

First experimental results on $t/t_{w}$ scaling of the
thermoremanent magnetization (TRM) were obtained soon after the
aging effects had been discovered. They demonstrated that there
are small but systematic deviations from full $t/t_{w}$ scaling in
the aging regime.\cite{oci85,alb87} It was shown that the TRM
curves for different waiting times can be successfully
superimposed if a power of the waiting time, $t_{w}^{\mu}$, with
$\mu<1$, is used in the analysis instead of the actual $t_{w}$.
This phenomenon is often referred to as ``subaging''. The apparent
age of the spin-glass state, determined from the scaling of the
TRM curves, increases slower than the waiting time $t_{w}$. It was
also shown \cite{oci85,alb87} that the quasiequilibrium decay at
short observation times should be taken into account properly in
order to achieve good scaling over a wide time range. However,
full $t/t_{w}$ scaling in the aging regime has never been derived
from experimental data. Deviations of $\mu$ from unity are small,
but persistent. The physical meaning of this scaling parameter
remains unclear even though different interpretations have been
proposed.\cite{bou94,rin00} All this suggests that there may be
some additional reasons for the lack of scaling, which have not
been taken into account yet.

On the theoretical side, the situation is more ambiguous.
Phenomenological phase-space models,\cite{bou92,bou95} describing
aging as thermally activated hopping over free-energy barriers,
suggest full $t/t_{w}$ scaling in the absence of finite-size
effects. In the mean-field dynamics, however, two different cases
are distinguished.\cite{bou97} Those mean-field models, where one
level of replica-symmetry breaking is exact (like the spherical
$p$-spin model), have only one time scale in the aging regime, and
$t/t_{w}$ scaling is predicted.\cite{cug93} Models with continuous
replica-symmetry breaking (like the Sherrington-Kirkpatrick
model), are characterized by an infinite number of time
scales.\cite{cug94} If decay of the correlation function in the
aging regime is viewed as a sequence of infinitesimal steps, then
each step takes much longer than the previous one.\cite{cug99}
Therefore, no $t/t_{w}$ scaling is expected in these models. It is
suggested\cite{bou97,cug99} that full $t/t_{w}$ scaling for all
times in the aging regime would rule out the continuous
replica-symmetry breaking scenario. Thus, a reliable experimental
evidence of presence (or absence) of $t/t_{w}$ scaling may help
determine validity of different theoretical pictures.

Numerical studies of spin-glass dynamics reveal an interesting
fact. The short-range 3-dimensional Edwards-Anderson (EA) model
exhibits good $t/t_{w}$ scaling,\cite{kis96,rit01} while there is
no such scaling in the infinite-range Sherrington-Kirkpatrick (SK)
model.\cite{mar98,tak97,rit94} However, validity of the EA model
for description of real spin-glasses has been questioned. Recent
simulations \cite{pic01} have shown no trace of rejuvenation
effects, observed in temperature-change experiments.\cite{jon00}
It was argued that either time scales, reached in simulations, are
too short, or the Edwards-Anderson model is not a good model for
real spin glasses.\cite{pic01} Further studies of both $t/t_{w}$
scaling and temperature-variation effects are needed to find out
which model is better suited for theoretical description of
spin-glass dynamics.

The present paper is devoted to a study of departures from full
$t/t_{w}$ scaling in spin-glass aging experiments. We hypothesize
that one of possible reasons for the lack of such scaling is the
influence of the cooling process. Our line of argument is the
following. A typical spin-glass relaxation experiment includes
cooling from above the glass temperature.\cite{nor87} Approach of
the measurement temperature is necessarily slow with possible
oscillations, because temperature has to be stabilized.
Temperature-cycling experiments,\cite{ref87,led91,ham92} as well
as aging experiments with different cooling rates,\cite{nor00}
have shown that thermal history near the measurement temperature
has a profound effect on the subsequent spin-glass behavior. Any
aging experiment is, therefore, a temperature-variation
experiment, and measured relaxation properties are determined by
both cooling and waiting time effects. As the waiting time
increases, the influence of the initial condition, set by the
cooling process, diminishes. This may lead to a systematic
deviation from full $t/t_{w}$ scaling, compatible with the
experimentally observable behavior. In order to extract relaxation
curves that exhibit good $t/t_{w}$ scaling, one has to average
experimental results over different initial conditions.

The paper is organized as follows. The next Section discusses
general properties of spin-glass relaxation and methods for
analysis of $t/t_{w}$ scaling. In Sec.~III.A, features of the
cooling process and properties of the TRM decay for $t_{w}=0$ are
studied. Sec.~III.B discusses two different types of deviations
from $t/t_{w}$ scaling. In Sec.~III.C, properties of the TRM and
the IRM are analyzed in detail. The effect of initial conditions
on scaling properties of measured relaxation is discussed in
Sec.~III.D. The last Section summarizes our arguments.

\section{\label{theor} Theoretical background}

\subsection{\label{theA} Linear response in spin glasses}

In a typical TRM experiment, a spin-glass sample is cooled down
from above the glass temperature $T_{g}$ to a measurement
temperature $T<T_{g}$ in the presence of a small magnetic field
$h$. It is then kept at the measurement temperature for some time
$t_{w}$, which is called waiting time. After that, the field $h$
is cut to zero, and a decay of the thermoremanent magnetization
(TRM) is measured as a function of observation time $t$, elapsed
after the field change. This decay depends on the waiting time
$t_{w}$ -- a phenomenon called aging. The total age of the system
is $t+t_{w}$. In a more complex IRM protocol, the sample is cooled
down at zero magnetic field and kept for a waiting time $t_{w1}$.
Then a small magnetic field is turned on, and, after an additional
waiting time $t_{w2}$, is turned off again. A subsequent decay of
the isothermal remanent magnetization (IRM) is observed. It
depends on both waiting times.

In order to describe time evolution of the system, the
autocorrelation function, $C(t_{1},t_{2})$, is introduced:
\begin{equation}
C(t_{1},t_{2})=(1/N)\sum_{i=1}^{N} \langle
S_{i}(t_{1})S_{i}(t_{2}) \rangle~~. \label{cor}
\end{equation}
It depends on two times, $t_{1}$ and $t_{2}$, measured from the
end of the cooling process. Response of the system at time $t_{1}$
to an instantaneous field $h$, present at time $t_{2}$, is
described by the response function:
\begin{equation}
R(t_{1},t_{2})=(1/N)\sum_{i=1}^{N} \delta \langle S_{i}(t_{1})
\rangle / \delta h(t_{2})~~. \label{res}
\end{equation}
In thermal equilibrium, both functions depend only on the time
difference, $t_{1}-t_{2}$, and they are related by the
fluctuation-dissipation theorem. If the system is out of
equilibrium, a generalization of this theorem is expected to
hold:\cite{bou97}
\begin{equation}
R(t_{1},t_{2})= \beta X[C] \, \partial C(t_{1},t_{2}) /
\partial t_{2}~~. \label{fdt}
\end{equation}
Here $\beta=1/k_{B}T$, and $X[C]$ is the fluctuation-dissipation
ratio, which is equal to unity in equilibrium. It is suggested
\cite{bou97} that, for long waiting times, $X$ depends on its time
arguments only through the correlation function, i.e.
$X(t_{1},t_{2})=X[C(t_{1},t_{2})]$.

Magnetic susceptibility, measured in spin-glass experiments, is
the integrated response. If the field is present during the
waiting time, $t_{w}$, the susceptibility, measured at the
observation time, $t$, is given by the following expression:
\begin{equation}
\chi(t+t_{w},t_{w})=\int_{0}^{t_{w}} R(t+t_{w},t') dt'~~.
\label{chi}
\end{equation}
Using Eq.~(3) for the response function and introducing a function
$Y[C]$ through a relation $\beta X[C]=dY/dC$,\cite{bou95} one
obtains the formula:
\begin{equation}
\chi(t+t_{w},t_{w})=Y[C(t+t_{w},t_{w})]-Y[C(t+t_{w},0)]~~.
\label{dif}
\end{equation}
The first term on the right is identified with the
$t_{w}$-dependent TRM susceptibility.\cite{bou95} The second term
corresponds to the TRM decay after zero waiting time. To avoid
confusion, we shall refer to it as ZTRM, and drop the second
argument. The function on the left is the IRM susceptibility for
$t_{w1}=0$ and $t_{w2}=t_{w}$. In what follows, the IRM will
always denote the isothermal remanent magnetization with
$t_{w1}=0$. It is characterized by the same waiting time, $t_{w}$,
as the TRM. We shall also use a corresponding observation time
instead of a total age as the first argument of each function. For
the ZTRM, the total age is equal to the observation time. Thus,
Eq.~(5) can be rewritten in the following form:
\begin{equation}
\chi_{IRM}(t,t_{w})=\chi_{TRM}(t,t_{w})-\chi_{ZTRM}(t+t_{w})~~.
\label{irm}
\end{equation}
The last formula expresses the well-known principle of
superposition, which has been verified experimentally in the case
of spin glasses.\cite{lun86}

It follows from Eq.~(6) that the TRM relaxation is a superposition
of two decays. The IRM is a response, associated with the waiting
time. The ZTRM is a response, related to the cooling process.
However, the ZTRM itself cannot be treated as a linear response
because of the changing temperature. This means that the thermal
history cannot be taken into account by simple addition of the
cooling time to the waiting time.

If memory of the initial state is strong, there will be no
one-to-one correspondence between the $t_{w}$-dependent
correlation and $t_{w}$-dependent linear response. The TRM depends
on the correlation $C(t+t_{w},t_{w})$, but it includes response of
the initial state. The IRM is a linear response, associated with
the waiting time only, but it depends on the correlation with the
initial state. Therefore, neither the TRM, nor the IRM can be
called the "true" $t_{w}$-dependent response.

The mean-field theory of aging phenomena relies on the assumption
of weak long-term memory. According to this assumption, the system
responds to its past in an averaged way, and its memory of any
finite time interval in the past is weak.\cite{cug94} This implies
that the lower integration limit, $t'=0$, in Eq.~(4) can be
replaced by any finite $t'=t_{0}$, as long as both $t$ and $t_{w}$
go to infinity.\cite{cug94} Thus, there should be no difference
between the TRM and the IRM in the long-time limit. Within this
asymptotic approach, the IRM is sometimes referred to as TRM.
\cite{cug99,rit94}

The linear response theory predicts that the measured
susceptibility, Eq.~(4), depends on the waiting time $t_{w}$.
However, it gives no predictions regarding $t/t_{w}$ scaling.
Eq.~(6) suggests that the TRM and the IRM, though characterized by
the same waiting time, cannot have the same scaling properties.
This is because they differ by the ZTRM, which is a one-time
quantity without a characteristic time scale. Two exceptional
cases are possible. First, the ZTRM decays so rapidly, that the
last term in Eq.~(6) can be neglected for sufficiently large
$t_{w}$. This case is usually considered in theoretical
studies,\cite{vin96,bou95} which assume that memory of the initial
condition is lost after very long waiting times, i.e.
$C(t+t_{w},0) \rightarrow 0$ as $t_{w} \rightarrow
\infty$.\cite{fra95} Second, the ZTRM changes so slowly, that
$\chi_{ZTRM}(t+t_{w})$ for long enough $t_{w}$ can be treated as a
non-zero constant. This approach is similar to the treatment of
the field-cooled susceptibility, which is also a one-time
quantity. We show in Sec.~III.B that neither of these two
conditions holds in spin-glass experiments.

\subsection{\label{theB} Analysis of $t/t_{w}$ scaling}

Spin-glass dynamics is characterized by at least two time scales.
The microscopic attempt time, $\tau_{0}$, is associated with the
quasiequilibrium decay at short observation times. The waiting
time, $t_{w}$, determines properties of the nonequilibrium
relaxation at long times. Both types of spin-glass behavior have
to be taken into account when $t/t_{w}$ scaling of relaxation
curves is analyzed.

Numerical studies of the off-equilibrium dynamics in the 3D
Edwards-Anderson model\cite{kis96} suggest that the
autocorrelation function, $C(t,t_{w})$, can be well approximated
by a product of two functions:
\begin{equation}
C(t,t_{w}) \propto t^{-\alpha} \Phi(t/\,t_{w})~~. \label{kis}
\end{equation}
The waiting time independent factor $t^{-\alpha}$ represents the
slow (on the logarithmic scale) quasiequilibrium decay at $t \ll
t_{w}$. The function $\Phi(t/t_{w})$ (which is approximately
constant for $t \ll t_{w}$) describes the faster nonequilibrium
relaxation at longer observation times. Of course, both $\alpha$
and $\Phi(x)$ depend on temperature.

A similar multiplicative ansatz \cite{comment} has been
successfully used for scaling experimental relaxation curves:
\cite{oci85,alb87}
\begin{equation}
\chi_{TRM}(t,t_{w}) \propto t^{-\alpha} F(t_{e}/\,t_{w}^{\,
\mu})~~. \label{sig}
\end{equation}
Two features distinguish Eq.~(8) from Eq.~(7). First, an effective
time $t_{e}$ (usually denoted by $\lambda$) is introduced. The age
of the system increases with the observation time as $t_{w}+t$,
and the time/age ratio decreases. In order to allow description of
the relaxation by the same age $t_{w}$, the effective time should
increase slower than the observation time $t$. This means that
$dt_{e}/t_{w}=dt/(t+t_{w})$ and $t_{e}=t_{w}\ln(1+t/t_{w})$.
Second, possible deviations from full $t/t_{w}$ scaling in the
aging regime are taken into account by the parameter $\mu$. In
this case, $dt_{e}/t_{w}^{\, \mu}=dt/(t+t_{w})^{\mu}$, and the
effective time is
\begin{equation}
t_{e}=t_{w}[(1+t/t_{w})^{1-\mu}-1]/(1-\mu)~~. \label{lam}
\end{equation}
At short observation times, $t \ll t_{w}$, the effective time is
equivalent to the observation time: $t_{e} \approx t$.

The $\mu$-scaling approach is very useful for studying departures
from $t/t_{w}$ scaling, no matter whether $\mu$ itself has a clear
physical meaning or not. We use this method in the present paper
and determine $\mu$ from juxtaposition of relaxation curves,
plotted versus $t_{e}/t_{w}^{\, \mu}$.

A different method for separating the quasiequilibrium and aging
regimes has also been successfully employed.\cite{vin96} It is
inspired by dynamical solution of mean-field models.
\cite{bou97,cug93,cug94} The following additive representation is
considered: \cite{vin96}
\begin{equation}
\chi_{TRM}(t,t_{w}) \propto
A(t/\tau_{0})^{-\alpha}+f(t_{e}/\,t_{w}^{\, \mu})~~. \label{sum}
\end{equation}
This approach yields better $t/t_{w}$ scaling ($\mu$ closer to
unity) than the previously discussed method. \cite{oci85,alb87} It
should be noted, however, that derivation of the last
formula\cite{vin96} employs an assumption that $C(t+t_{w},0)=0$.
Therefore, Eq.~(10) is exact only asymptotically (large $t_{w}$),
when long-term memory is weak. Nevertheless, this method is
justified, because it gives good results on experimental time
scales.

\section{\label{exper}Experimental results and analysis}

The purpose of this paper is to study how different initial
conditions affect aging phenomena in spin-glass experiments. By
the initial condition we mean history of the spin-glass state
before the waiting time begins. We do not attempt to prove
existence or absence of full $t/t_{w}$ scaling. We try to
understand, on the qualitative level, where observable departures
from the full scaling in the aging regime come from.

All experiments were performed on a single crystal of Cu:Mn 1.5 at
\%, a typical Heisenberg spin glass with a glass temperature of
about $15.2~K$. The single crystal was used to avoid possible
complications due to finite-size effects. Another advantage of
this sample is its high de Almeida-Thouless (AT) critical line.
\cite{zot01} For example, the AT field at $T/T_{g}=0.87$, the
highest measurement temperature in our experiments, is about
$600~Oe$. This enables us to use a relatively large field change,
$\Delta H=10~Oe$, and still work well within the linear response
regime. A commercial Quantum Design SQUID magnetometer was used
for all the measurements. This equipment has been optimized for
precision and reproducibility, rather than speed. Thus, cooling is
relatively slow, and this allows us to study cooling effects in
detail.

\subsection{\label{expA} Cooling process and ZTRM}

Spin-glass behavior is characterized by memory effects: properties
of measured spin-glass relaxation depend on history of a sample
below $T_{g}$. The cooling process is an integral part of this
history. It has been shown that, in real spin glasses, details of
the experimental protocol well above the measurement temperature,
$T$, do not affect the measured relaxation.\cite{ref87} However,
the thermal history in the immediate vicinity of the measurement
temperature (say, $\delta T < 0.5~K$) has a strong impact on the
observed spin-glass behavior. This fact suggests that the cooling
and waiting time effects cannot be separated and should be studied
simultaneously.

A typical cooling protocol is exhibited in the inset of Fig.~1.
The temperature drops rapidly to well below the measurement
temperature, than rises to $T+\delta T$, and slowly approaches $T$
from above. By definition, the experimental time starts when the
measurement temperature, $T$, is finally reached.

The approach of the measurement temperature is shown in the main
body of Fig.~1. For relatively high temperatures, $\delta T
\approx 0.1~K$. For the lowest temperature, however, $\delta T
\approx 0.3~K$. Comparison with results of temperature-cycling
experiments\cite{ref87} suggests that the initial undercool is not
very important, because of its large (several Kelvin) magnitude.
The subsequent overshoot, however, may be expected to play a
significant role in spin-glass dynamics, and cannot be neglected
in our experiments. It should be noted that Fig.~1 exhibits the
temperature curves for helium gas, used as a heat transfer agent
within a sample chamber. The actual temperature of the spin-glass
sample lags behind. This means that, in reality, the undercool is
probably less pronounced, and the sample spends more time above
the measurement temperature.

\begin{figure}
\resizebox{\columnwidth}{!}{\includegraphics{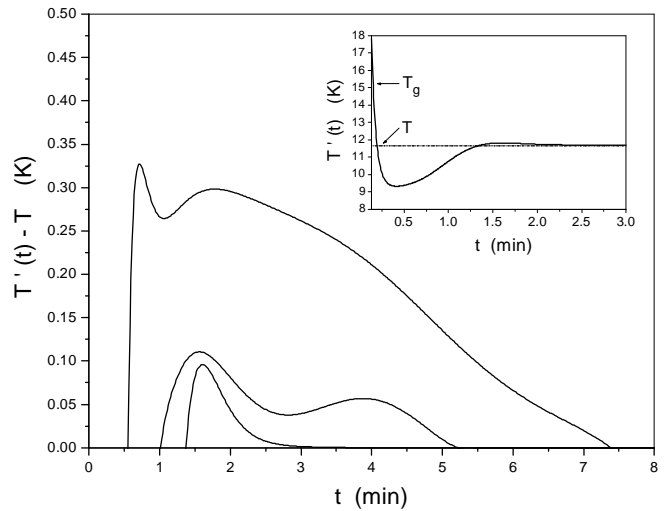}}
\caption{\label{F1} Approach of the measurement temperature, $T$,
during the cooling process. The temperature curves $T'(t)$, from
top to bottom, are for $T/T_{g}=0.3, 0.4$, and $0.77$,
respectively. The inset shows the whole cooling process for
$T/T_{g}=0.77$.}
\end{figure}

It is instructive to compare the cooling procedure in Fig.~1 with
an experimental protocol including a positive temperature
cycle.\cite{ref87} In this protocol, the sample is field-cooled to
the measurement temperature, $T$, and kept for a long waiting time
$t_{w1}$. Then it is heated up to $T+ \delta T$ and annealed for a
short time $t_{w2}$. After that, it is cooled down and kept at the
initial temperature $T$ for additional time $t_{w3}$. Then the
field is cut to zero, and the TRM relaxation is measured. It turns
out that this relaxation follows a reference curve with
$t_{w}=t_{w3}$ at short observation times, then breaks away and
moves towards a reference curve with $t_{w}=t_{w1}$. Therefore,
the spin-glass state initially exhibits memory of only those
events that happened after the temperature cycle. But, as time
progresses, it begins to recall its earlier history. Another
relevant analogy is an experimental protocol with a negative
temperature shift after initial waiting.\cite{led91} In both
experiments, the relaxation curve after the temperature change
cannot be merged with any of the reference curves measured at a
constant temperature.

The main idea of this paper is that any spin-glass aging
experiment is, at the same time, a temperature-cycling experiment.
All effects, which manifest themselves in temperature-cycling
experiments, also appear in aging experiments. Our reasoning is
based on the hierarchical phase-space picture of spin-glass
dynamics.\cite{ref87,led91} Let us consider a typical cooling
process and imagine that the system spends some short time at
temperature $T+\delta T$. Several metastable states, separated by
free-energy barriers, become populated during that time. As
temperature is lowered, these states split in a hierarchical
fashion and produce new metastable states with new barriers. The
old barriers grow steeply, and, for any value of temperature,
there exist barriers, diverging at this temperature.\cite{ham92}
Therefore, when the measurement temperature $T$ is finally
reached, there is a large number of metastable states, separated
by barriers of all heights. In other words, the cooling process
creates an initial state with a broad spectrum of relaxation
times. This situation corresponds to ``existence of domains of all
sizes within the initial condition'' in the real-space
picture.\cite{jon00}

If a waiting time, $t_{w}$, follows the cooling, properties of the
subsequent relaxation will be similar to those in the
positive-temperature-cycle experiment. At short observation times,
the relaxation will be governed by $t_{w}$. At longer observation
times, it will break away and slow down, because the long-time
metastable states with high barriers, created by the cooling
process, will come into play. This effect will be more pronounced
after shorter waiting times.

It is important to mention that the results of the
temperature-variation experiments can be explained within the
phenomenological real-space picture, if a strong separation of
time and length scales is postulated.\cite{bou02} The experimental
data,\cite{ham92} that suggest divergence of free-energy barriers
at any temperature below $T_{g}$, can be reanalyzed in terms of
barriers, which vanish at $T_{g}$, and remain finite at lower
temperatures.\cite{bou02} The barriers, nevertheless, grow with
decreasing temperature. This fact is sufficient for understanding
results of the present paper.

Properties of the initial state, produced by the cooling process,
can be studied by measuring a TRM decay for $t_{w}=0$, which we
call ZTRM. The sample is cooled down in the presence of a small
field, $H=10~Oe$. When the temperature is stabilized, the field is
cut to zero, and decay of the ZTRM is recorded.

Fig.~2 shows the logarithmic relaxation rates of this decay,
corresponding to different temperatures. Each curve has a peak,
and we call a position of this peak the ``effective cooling
time'', $t_{c}^{e\!f\!f}$. Each magnetization curve is normalized
by one at $t=t_{c}^{e\!f\!f}$ before differentiation is performed.
It can be seen from Fig.~2 that the peaks become much broader as
temperature decreases. This means that the spectrum of relaxation
times broadens as well, and it is not dominated by a single time
scale.

\begin{figure}
\resizebox{\columnwidth}{!}{\includegraphics{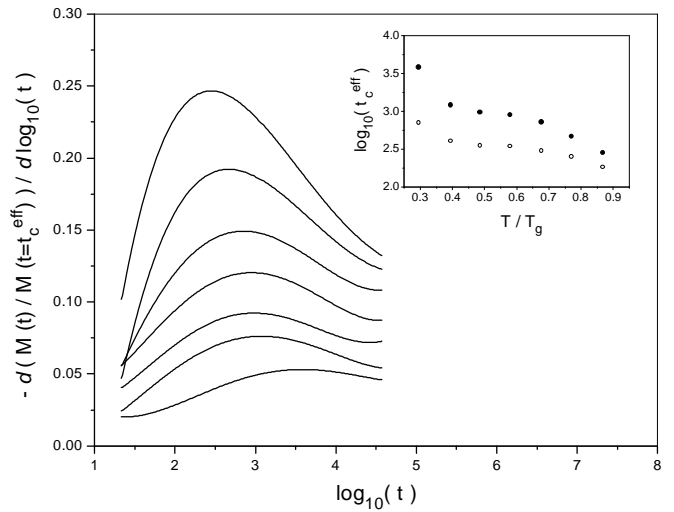}}
\caption{\label{F2} The logarithmic relaxation rate for the ZTRM.
The measurement temperatures, from top to bottom, are
$T/T_{g}=0.87, 0.77, 0.68, 0.58, 0.5, 0.4$, and $0.3$. The inset
shows temperature dependence of logarithms of the effective
cooling time, $t_{c}^{e\!f\!f}$ (solid symbols), and the effective
scaling time, $t_{s}^{e\!f\!f}$ (open symbols).}
\end{figure}

Even though the effective cooling time, $t_{c}^{e\!f\!f}$,
corresponds to the maximum of the relaxation rate, it cannot be
treated as a regular waiting time $t_{w}$. We have tried to scale
each ZTRM curve with the relaxation curves for longer waiting
times, using the $\mu$-scaling approach and taking
$t_{c}^{e\!f\!f}$ as the scaling parameter. No value of $\mu$ can
give even approximate scaling, especially at short times. It turns
out, however, that the ZTRM curve can be merged with the other TRM
curves (using the same $\mu$ as used for the longer waiting
times), if a much shorter characteristic time for the ZTRM is
introduced. We call this time the ``effective scaling time'',
$t_{s}^{e\!f\!f}$. It is several times shorter than the effective
cooling time. The inset of Fig.~2 exhibits logarithms of
$t_{c}^{e\!f\!f}$ and $t_{s}^{e\!f\!f}$ for different
temperatures. The error bars for the logarithms are typically 0.1,
but about 0.2 for the lowest temperature.

The fact that one can consider at least two characteristic times
for the ZTRM agrees with what is expected from relaxation in the
positive-temperature-cycle experiment. The spin-glass behavior at
short observation times is governed by $t_{s}^{e\!f\!f}$, which
can be associated with ordinary aging effects during the slow
approach of the measurement temperature. However, the long-time
relaxation is strongly influenced by the metastable states,
separated by high barriers, resulting from the temperature change.
Thus, the relaxation rate peaks at $t_{c}^{e\!f\!f}$, and not at
$t_{s}^{e\!f\!f}$. This argument can be generalized to include the
TRM experiments with finite waiting times. The logarithmic
relaxation rate has a peak at $t_{w}^{e\!f\!f}$, the effective
waiting time. It is well known that $t_{w}^{e\!f\!f}$ for the TRM
is greater than $t_{w}$.\cite{nor00} We believe that this shift in
the maximum of the relaxation rate is caused by those long-time
metastable states, which are created during the cooling process.

It is important to note that memory of the initial state depends
not only on the cooling protocol, but also on the overall
complexity of the free-energy landscape. This can be illustrated
by measurements of the ZTRM in the presence of a high constant
field. The minimum possible overlap, $q_{min}(H)$, between two
states increases with increasing magnetic field, $H$.\cite{par83}
Therefore, certain constraints on barrier heights are imposed by
the field, leading to a faster relaxation. The experiment is
performed in the following way. First, the sample is cooled down
to the measurement temperature at the field $H+\Delta H$. The
field is changed to $H$, and the zero waiting time magnetization
(ZTRM) is measured for some time. Second, the sample is warmed up
to above $T_{g}$, and cooled down again, all in the presence of
the same field $H$. The field-cooled magnetization (MFC) is
measured for the same time. The difference of these two
magnetizations is then fitted to a power law at long observation
times: $ZTRM(t,H,\Delta H)-MFC(t,H) \propto t^{-\lambda (H)}$. The
values of $H$ were varied from $50~Oe$ to $1000~Oe$. The field
change was small, $\Delta H = 10~Oe$, so that the response was
always linear.

Fig.~3 exhibits the measured ZTRM and MFC decays for $H=500~Oe$.
One can see that time dependence of the MFC cannot be neglected at
high fields. The inset shows the relaxation exponent $\lambda
(H)$. The exponent appears to be a linear function of $q_{min}
\propto H^{2/3}$. This is not surprising, because its temperature
dependence is also close to linear, and the AT critical line,
$T_{g}-T_{c}(H) \propto H^{2/3}$, has a profound effect on
spin-glass dynamics.\cite{zot01} The estimated value of $\lambda
(H)$ near the AT line, which corresponds to $H_{AT} \sim 1400~Oe$
at this temperature, is $0.27 \pm 0.03$. It is interesting that
this number compares well with the Monte Carlo result, $\lambda
\approx 0.25$, for the 3D EA model at zero magnetic
field.\cite{kis96}

\begin{figure}
\resizebox{\columnwidth}{!}{\includegraphics{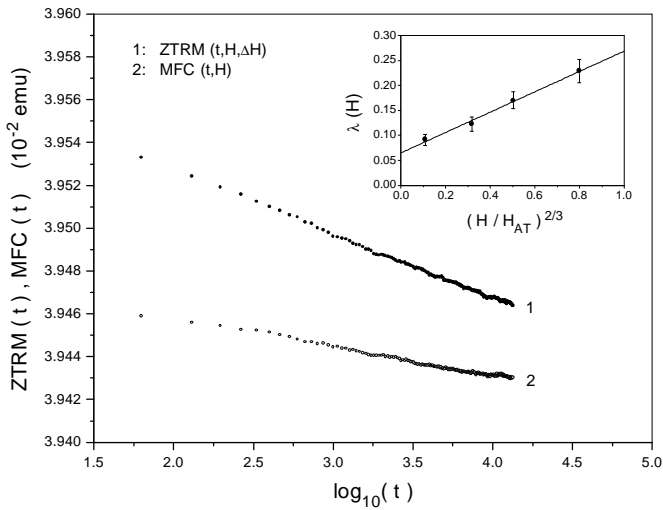}}
\caption{\label{F3} Time dependences of the ZTRM and MFC, measured
at $T/T_{g}=0.79$ and $H=500~Oe$. The field change is $\Delta
H=10~Oe$. The inset shows the relaxation exponent, $\lambda (H)$,
as a function of $H^{2/3}$.}
\end{figure}

The observed increase in $\lambda (H)$ suggests that energy
barriers, created by the cooling process, are lower, if cooling is
performed in the presence of a high field. This happens because
the accessible phase space is limited by $q_{min}(H)$. The
low-field initial state in our experiments appears to be more
complex, i.e. characterized by higher barriers, than the initial
state in Monte Carlo simulations.

\subsection{\label{expB} Subaging or superaging?}

It was shown in Sec.~II.A that the TRM relaxation can be
decomposed into a sum of two decays: the IRM, which is a response
of the system associated with the waiting time, $t_{w}$, and the
ZTRM, which is a response of the initial state.

Fig.~4 displays the experimental TRM and ZTRM curves, and their
difference, for $T/T_{g}=0.87$ and $t_{w}=1000~s$. One can see
that the response of the initial state dominates the measured TRM
relaxation. This effect is particularly pronounced at low
temperatures, where thermally activated processes are slow. Fig.~4
suggests that the ZTRM cannot be neglected, nor can it be regarded
as a constant. Therefore, the TRM and IRM relaxation curves will
have different scaling properties, and should be studied
simultaneously.

\begin{figure}
\resizebox{\columnwidth}{!}{\includegraphics{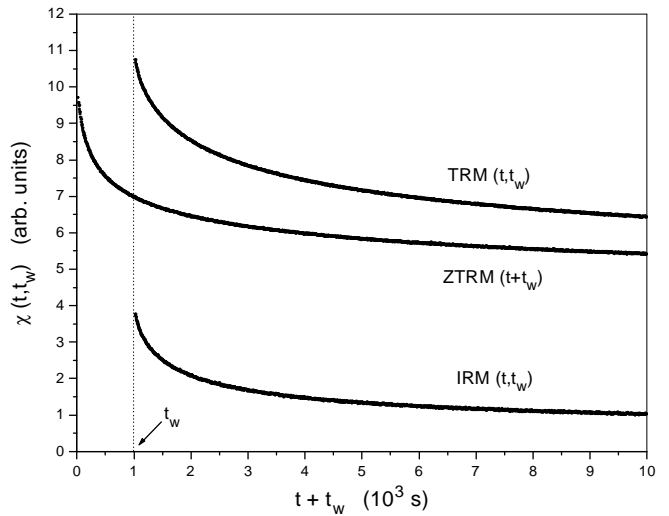}}
\caption{\label{F4} Experimental time dependence of the
$\chi_{TRM}(t,t_{w})$ and $\chi_{IRM}(t,t_{w})$ for $T/T_{g}=0.87$
and $t_{w}=1000~s$. The TRM relaxation for zero waiting time is
referred to as ZTRM. Note its large magnitude in comparison with
the TRM for $t_{w} \neq 0$.}
\end{figure}

In this paper, we consider a multiplicative representation for the
measured relaxation curves, in analogy with Eqs.~(7) and (8):
\begin{equation}
\chi_{TRM}=\phi(t/t_{w}) \, \rho_{TRM}(t,t_{w},t_{c})~~;
\label{ftr}
\end{equation}
\begin{equation}
\chi_{IRM}=\phi(t/t_{w}) \, \rho_{IRM}(t,t_{w},t_{c})~~.
\label{fir}
\end{equation}
Here we made an assumption that both the TRM and the IRM are
characterized by the same scaling part, $\phi(t/t_{w})$. This is
certainly correct in the limit of very long waiting times, when
the two functions are essentially equal. Analysis of experimental
data, based on Eqs.~(11) and (12), suggests that this approach
works well on the real time scales.

Systematic departures from full $t/t_{w}$ scaling are described by
the factors $\rho_{TRM}$ and $\rho_{IRM}$. If violations of
$t/t_{w}$ scaling were only due to the quasiequilibrium behavior,
these factors would depend only on the observation time, $t$. We
assume, however, that there is an interplay between the cooling
and waiting-time effects. Thus, we include the waiting time,
$t_{w}$, and an additional argument $t_{c}$, which indicates
dependence on the cooling process.

The correlation function, $C(t+t_{w},t_{w})$, defined by Eq.~(1),
is independent of $t_{w}$ at $t=0$. Numerical studies of the SK
model demonstrate \cite{mar98,tak97} that the correlation curves
for different waiting times, plotted versus $t/t_{w}$, cross at
one point, corresponding to $t \approx t_{w}$. This is not
surprising, because transition from the quasiequilibrium to the
aging regime occurs at $t \approx t_{w}$, and the correlation at
this point is near $q_{EA}$. If the linear response susceptibility
depends on its time arguments only through the correlation
function, i.e. $\chi=\chi(C)$, the susceptibility curves should
also cross at one point. Experimental results do not exhibit this
property. The TRM curves for long waiting times lie below the
curves for short waiting times, when plotted vs. $t/t_{w}$. The
IRM curves demonstrate the opposite behavior.

In the present paper, we normalize both the TRM and the IRM decays
by one at $t=t_{w}$, and treat them as ``normal'' relaxation
functions. This approach has several advantages. First, all
departures from full $t/t_{w}$ scaling, reflected in the shapes of
relaxation curves, can be observed clearly. Second, the shapes of
TRM and IRM decays can be compared in detail, regardless of the
fact that their magnitudes are different. Third, experimental
results can be directly compared with results of numerical
simulations. We believe that this approach is physically justified
because the influence of the cooling process leads to systematic
changes in the \emph{shapes} of relaxation curves, as discussed in
Sec.~III.A. This method is different from the one, traditionally
used to study $t/t_{w}$ scaling.\cite{vin96,oci85} It provides new
insights into aging behavior of real spin glasses.

Fig.~5 exhibits the TRM and IRM relaxation curves, measured at
$T/T_{g}=0.87$, for two waiting times, $t_{w}=1000~s$ and
$t_{w}=6310~s$. The TRM data demonstrate the familiar subaging
pattern: the relaxation, plotted versus $t/t_{w}$, is faster for
the longer waiting time. The IRM results show a quite different
behavior: the relaxation as a function of $t/t_{w}$ slows down as
$t_{w}$ increases. This effect is called ``superaging''.
\cite{bou97}

\begin{figure}
\resizebox{\columnwidth}{!}{\includegraphics{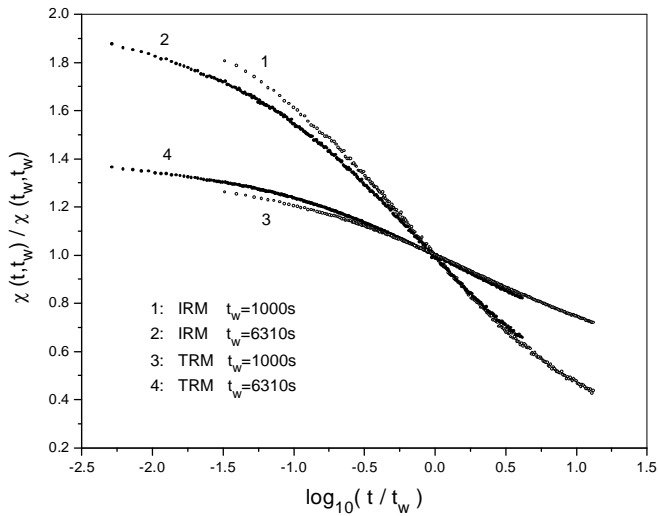}}
\caption{\label{F5} Relaxation curves $\chi_{IRM}(t,t_{w})$ and
$\chi_{TRM}(t,t_{w})$ for two waiting times, normalized by one at
$t=t_{w}$. The temperature is $T/T_{g}=0.87$. The IRM decay as a
function of $t/t_{w}$ is slower for longer waiting times
(``superaging''). The TRM decay is faster for longer waiting times
(``subaging'').}
\end{figure}

The phenomenon of superaging may seem rather unusual from an
experimentalist's point of view, because it has never been
observed in TRM experiments. It turns out, however, that this is a
predominant effect in Monte Carlo simulations. Fig.~4 in Marinari
\emph{et al.},\cite{mar98} Fig.~4 in Takayama \emph{et
al.},\cite{tak97} and Fig.~4 in Cugliandolo \emph{et
al.}\cite{rit94} clearly demonstrate the superaging behavior. In
all these simulations of the SK model, the correlation function,
$C(t+t_{w},t_{w})$, plotted versus $t/t_{w}$, decays slower for
longer waiting times. In the case of the EA model, it is more
difficult to distinguish between the subaging and superaging
effects, because the correlation curves exhibit good $t/t_{w}$
scaling in the aging regime. However, the relaxation exponent
$\lambda(T,t_{w})$, defined as $C(t+t_{w},t_{w}) \propto
t^{-\lambda}$ for $t \gg t_{w}$, decreases slightly with
increasing $t_{w}$, according to Fig.~2 in Kisker \emph{et
al.}\cite{kis96} This fact suggests that the superaging behavior
appears in the EA model as well.

\begin{figure}
\resizebox{\columnwidth}{!}{\includegraphics{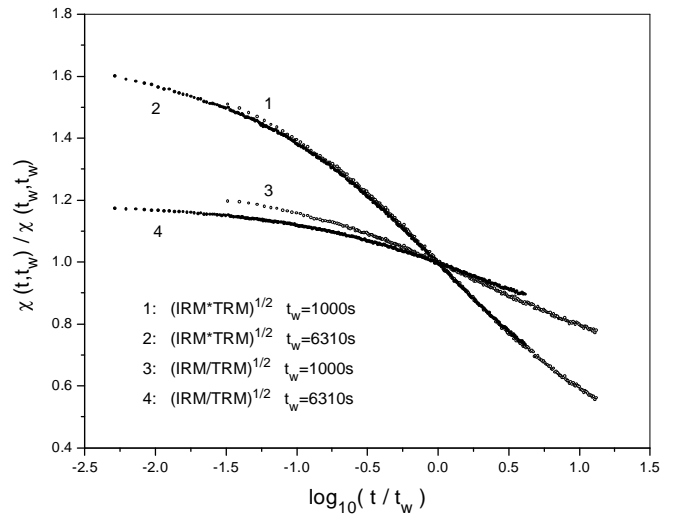}}
\caption{\label{F6} Factorization of the relaxation curves from
Fig.~5. $f=(\chi_{IRM} \, \chi_{TRM})^{1/2}$ is an average over
two different initial conditions, which shows systematic
improvement in $t/t_{w}$ scaling.
$g=(\chi_{IRM}/\chi_{TRM})^{1/2}$ describes departures from the
average and does not scale well.}
\end{figure}

Because the experimental TRM and IRM curves have the opposite
scaling properties (subaging versus superaging), the functions
$\rho_{TRM}$ and $\rho_{IRM}$, describing departures from full
$t/t_{w}$ scaling in Eqs.~(11) and (12), may be expected to have
inverse effects on the function $\phi(t/t_{w})$. Following this
observation, we define two auxiliary functions, $f$ and $g$. The
function $f$ is a geometric mean of $\chi_{TRM}$ and $\chi_{IRM}$,
and the function $g$ describes departures from this mean:
\begin{equation}
f(t,t_{w},t_{c})=(\chi_{IRM} \, \chi_{TRM})^{1/2}~~; \label{f}
\end{equation}
\begin{equation}
g(t,t_{w},t_{c})=(\chi_{IRM} / \chi_{TRM})^{1/2}~~. \label{g}
\end{equation}
With this definition of $f$, the scaling part, $\phi(t/t_{w})$, is
left unchanged, while deviations from full scaling are averaged
out, at least partially. Physically, the function $f$ represents
an average over two different initial conditions. Of course, if
both $\chi_{TRM}$ and $\chi_{IRM}$ are normalized by one at
$t=t_{w}$, their geometric mean is close to the arithmetic mean.
The IRM and TRM relaxation curves can be factorized as follows:
\begin{equation}
\chi_{IRM}=f \, g~~;~~~~~~~\chi_{TRM}=f/g~~. \label{fng}
\end{equation}

Fig.~6 exhibits the functions $f$ and $g$ for $T/T_{g}=0.87$. Both
of them demonstrate the superaging behavior, but the function $f$
shows much better $t/t_{w}$ scaling. The function $g$ is
responsible for the faster relaxation in the IRM experiments, and
the slower relaxation in the TRM experiments.

\subsection{\label{expC} Comparison of TRM and IRM}

In order to make the above arguments quantitative, we analyze the
$t/t_{w}$ scaling of relaxation curves using the $\mu$-scaling
approach, discussed in Sec.~II.B. We optimize $\mu$ to achieve the
best possible scaling over a wide range of observation times, with
particular attention to the aging regime, $t>t_{w}$. Inclusion of
the quasiequilibrium decay with the exponent $\alpha$ improves
scaling at short times, but it does not help at longer times. In
the present analysis, the quasiequilibrium behavior is not taken
into account explicitly. This gives values of $\mu$ a little
further from unity, but does not affect any conclusions about
scaling in the aging regime.

Fig.~7 exhibits values of the parameter $\mu$ for the TRM, IRM,
and their combinations.

\begin{figure}
\resizebox{\columnwidth}{!}{\includegraphics{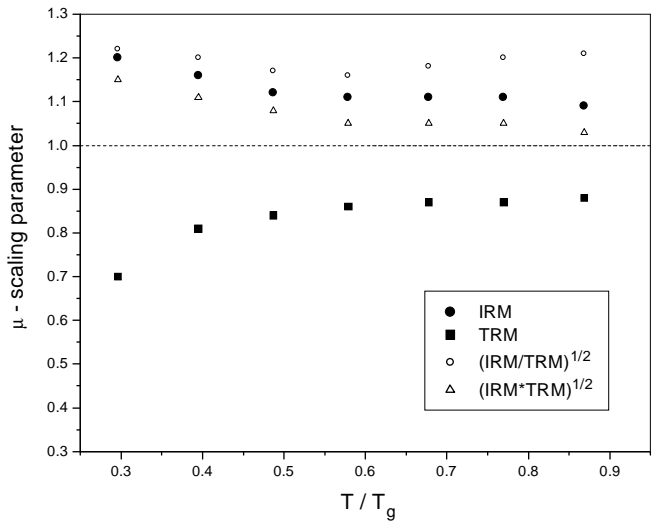}}
\caption{\label{F7} The scaling parameter $\mu$ as a function of
temperature for different relaxation functions. Note that $\mu <
1$ for the TRM (``subaging''), and $\mu > 1$ for the IRM
(``superaging'').\\}
\end{figure}

In the case of the TRM, $\mu$ lies between 0.8 and 0.9 for
relatively high temperatures, but drops visibly at low
temperatures, $T/T_{g} < 0.4$. Similar results have been obtained
for various spin-glass samples.\cite{vin96,oci85,alb87} The error
bars in our analysis are typically 0.02. At the lowest
temperature, however, they are about 0.1 for the TRM, and about
0.05 for the IRM.

The values of $\mu$ for the IRM are greater than unity. They are
near 1.1, but tend to increase at low temperatures. According to
Fig.~7, there is a certain symmetry with respect to $\mu=1$
between the values of $\mu$ for the TRM and the IRM. We have also
found that, in the case of the IRM, the parameter $\mu$ is very
sensitive to any imperfections of the scaling analysis. For
example, if the actual observation time is slightly longer, than
the time $t$, used in the analysis, the corresponding value of
$\mu$ will seem closer to unity. This issue is important when a
superconducting magnet is used, because it takes some time to warm
up the persistent switch, and then to cool it down again. The
results for the TRM are less sensitive to these effects.

Fig.~7 also displays $\mu$ for the functions $f$ and $g$, given by
Eqs.~(13) and (14). The value of $\mu$ for the function $f$ is
about 1.05, while for the function $g$ it is near 1.20. Thus, both
the TRM and the IRM relaxations can be presented as products of
two factors, according to Eq.~(15). The rapidly changing $f$
exhibits fairly good $t/t_{w}$ scaling. The relatively slow $g$
describes departures from this scaling. This function becomes
closer to a constant as $t_{w}$ increases, suggesting that both
the TRM and the IRM curves scale better for longer waiting times.
At low temperatures, however, the differences between $f$ and $g$
tend to diminish.

\begin{figure}
\resizebox{\columnwidth}{!}{\includegraphics{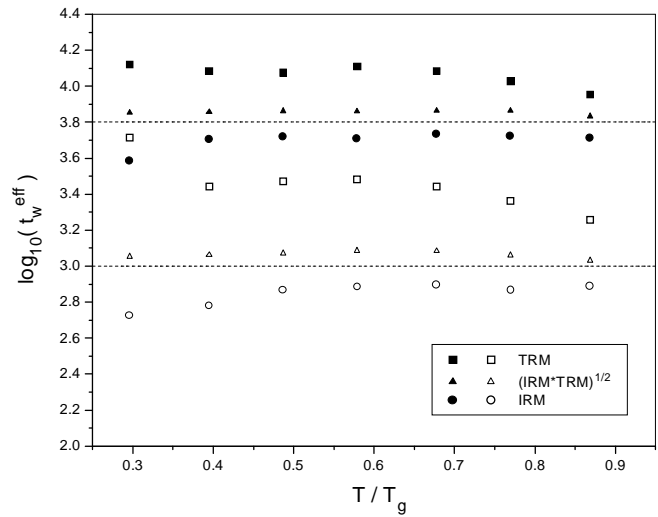}}
\caption{\label{F8} Logarithm of the effective waiting time,
$t_{w}^{e\!f\!f}$, as a function of temperature. The solid symbols
correspond to $t_{w}=6310~s$, and the open symbols -- to
$t_{w}=1000~s$. Note that $t_{w}^{e\!f\!f} > t_{w}$ for the TRM,
and $t_{w}^{e\!f\!f} < t_{w}$ for the IRM.\\}
\end{figure}

Fig.~8 exhibits values of logarithm of the effective waiting time,
$t_{w}^{e\!f\!f}$, for different relaxation functions. The
relaxation curves were fitted using a 5-order polynomial fit, and
then differentiated with respect to $\log_{10}(t)$. The effective
waiting time, $t_{w}^{e\!f\!f}$, corresponds to the maximum of
this derivative. The error bars for $\log_{10}(t_{w}^{e\!f\!f})$
are typically 0.05, but about 0.1 for the lowest temperature. It
can be seen from Fig.~8, that $t_{w}^{e\!f\!f} > t_{w}$ for the
TRM, but $t_{w}^{e\!f\!f} < t_{w}$ for the IRM. The values of the
effective waiting time for the function $f$ are greater than
$t_{w}$, but they all differ from $t_{w}$ by less than 3\%.
Therefore, the average of the TRM and the IRM gives better results
for the effective waiting time as well.

Fig.~9 provides a further insight into the nature of these
phenomena. It displays values of the relaxation exponent,
$\lambda(T,t_{w})$, defined through $\chi(t,t_{w}) \propto
t^{-\lambda(T,t_{w})}$ for $t \gg t_{w}$. This definition is based
on the fact that time dependence of spin-glass relaxation is
essentially algebraic in the aging regime.\cite{bou92,bou95,kis96}
The value of $\lambda(T,t_{w})$ is determined from the slope of
the linear fit to $\log_{10}(\chi)$ as a function of
$\log_{10}(t)$. Because of time limitations in our experiments,
the fitting was performed in the interval $t/t_{w}=1.5...10$ for
$t_{w}=1000~s$, and in the interval $t/t_{w}=1.5...5$ for
$t_{w}=6310~s$. In the case of the ZTRM, the effective cooling
time, $t_{c}^{e\!f\!f}$, was used instead of $t_{w}$. The
algebraic approximation works well, and the error bars for the
relaxation exponents are about 1\%.

\begin{figure}
\resizebox{\columnwidth}{!}{\includegraphics{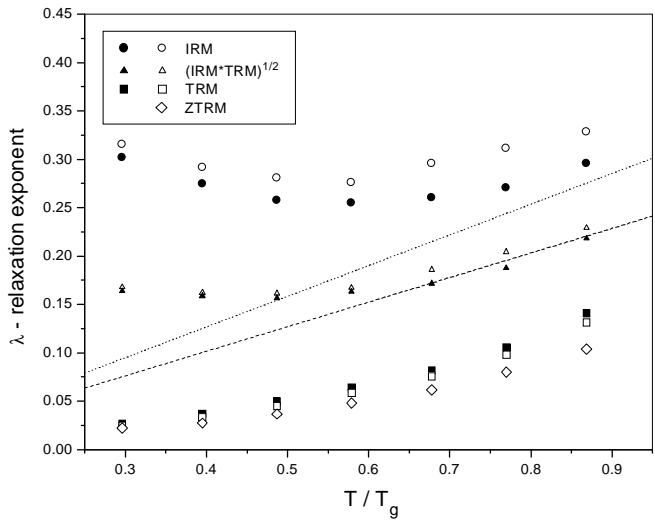}}
\caption{\label{F9} The relaxation exponent, $\lambda(T,t_{w})$,
as a function of temperature. The solid symbols correspond to
$t_{w}=6310~s$, and the open symbols -- to $t_{w}=1000~s$. The
open diamonds denote $\lambda$ for the experimental ZTRM. The
upper and lower straight lines are Monte Carlo results for
$t_{w}=0$ and $t_{w}=1000~mcs$, respectively (from Kisker \emph{et
al.}, Ref.~13).}
\end{figure}

The first conclusion one can draw from Fig.~9 is that the zero
waiting time decay, ZTRM, is the \emph{slowest} relaxation
measured in the aging regime. This result is in sharp
contradiction with results of numerical simulations, which show
that the zero waiting time decay is the \emph{fastest} possible
relaxation. The dotted (upper) line in Fig.~9 represents
$\lambda(T)$ for $t_{w}=0$ from the Monte Carlo studies of the EA
model by Kisker \emph{et al.}\cite{kis96} Our experimental values
of the $\lambda(T)$ for the ZTRM are about 3 times lower. We have
also tried to estimate the relaxation exponent for the SK model,
using numerical results of Marinari \emph{et al.}\cite{mar98} and
Takayama \emph{et al.}\cite{tak97}, and fitting them in the
interval $t/t_{w}=1...100$. In the first case, for $T/T_{g}=0.5$
and $t_{w}=8~mcs$, we find $\lambda(T,t_{w}) \approx 0.15$. In the
second case, for $T/T_{g}=0.4$ and $t_{w}=16~mcs$, we determine
$\lambda(T,t_{w}) \approx 0.20$. These numbers are similar to
those for the EA model.

Fig.~9 demonstrates that the TRM decay at $t>t_{w}$ becomes faster
as the waiting time increases, while the IRM relaxation becomes
slower. The corresponding values of $\lambda(T,t_{w})$ move
towards each other and towards the dashed (lower) line, which
represents numerical results for the EA model with
$t_{w}=1000~mcs$. \cite{kis96} One can see that the function $f$,
the average of the TRM and the IRM, has relaxation exponents,
which (at relatively high temperatures) are very close to those
from the Monte Carlo simulations. These exponents decrease
slightly with increasing $t_{w}$, as the numerical results do.

\subsection{\label{expD} A possible origin of $\mu$}

We are now in a position to offer a tentative explanation for
these phenomena. It is based on two observations. First, there is
a certain degree of symmetry between the TRM and the IRM with
respect to the ``ideal'' case, according to Figs.~7-9. Of course,
this symmetry is not exact and depends on a particular realization
of the cooling process. Second, when the waiting time, $t_{w}$, is
very large, the TRM and IRM become essentially equal, according to
Eq.~(6). This situation corresponds to the weak long-term
memory.\cite{cug94,bou97} These two observations suggest that the
experimental TRM and IRM decays approximate the true spin-glass
relaxation from the opposite sides.

Let us first consider a typical TRM experiment. When the cooling
is over, a number of metastable states, separated by free-energy
barriers of different heights, are already populated. We shall
refer to them loosely as the initial state. The response of this
state, the ZTRM, is slow and cannot be characterized by a single
well-defined time scale. The initial state is not random: it has
already evolved towards some equilibrium state, and this makes it
energetically favorable. During the waiting time, evolution
continues in the same direction. Thus, in the case of the TRM,
aging and cooling effects work together. The spin-glass state
appears to be older. The characteristic barrier is higher than
$\Delta(T,t_{w})=k_{B}T\ln(t_{w}/\tau_{0})$, the barrier,
associated with the waiting time $t_{w}$.\cite{led91} The TRM
decay is slower than the ideal $t_{w}$-dependent relaxation, which
can be characterized by some exponent $\lambda_{0}$. As the
waiting time increases, influence of the initial state diminishes.
The relaxation, plotted versus $t/t_{w}$, becomes faster, and its
dependence on $t_{w}$ -- stronger. This leads to the subaging
behavior. Therefore, for the TRM experiments, we would expect
$\mu<1$, $t_{w}^{eff}>t_{w}$, and $\lambda<\lambda_{0}$.

Let us now discuss a typical IRM experiment. In this case, a small
magnetic field is applied after the cooling process and before the
waiting time. The initial state, created by the cooling, is again
partially equilibrated. The subsequent field change makes it
energetically unfavorable. This is a consequence of chaotic nature
of the spin-glass state with respect to magnetic field. Even if
the field change is very small, the new equilibrium state (reached
at the end of the relaxation process) is very different from the
old one.\cite{par83} The system can no longer evolve in the same
direction. Moreover, it has to escape from the initial state.
Therefore, in the case of the IRM, aging and cooling effects work
against each other. During the same waiting time, $t_{w}$, the
system will have to overcome barriers, produced by the cooling
process, \emph{and} explore a different part of the phase space,
thus exhibiting aging. As a result, the spin-glass state looks
younger. The characteristic barrier, corresponding to the waiting
time, $t_{w}$, is now lower than $\Delta(T,t_{w})$ for the ideal
aging. The IRM decay is faster than the ideal relaxation. However,
memory of the initial state weakens as the waiting time increases.
The IRM relaxation as a function of $t/t_{w}$ becomes slower and
more dependent on $t_{w}$. This leads to the superaging behavior.
Therefore, the IRM experiments should yield $\mu>1$,
$t_{w}^{eff}<t_{w}$, and $\lambda>\lambda_{0}$.

If this explanation is reasonable, the observed difference in
scaling properties of the TRM and the IRM is a result of the
different initial conditions for aging. Departures from full
$t/t_{w}$ scaling are present in the aging regime not because
something goes wrong at long waiting times (e.g. the interrupted
aging\cite{bou94}), but because results for short waiting times
are strongly affected by the initial state. An average over
different initial conditions should improve $t/t_{w}$ scaling.

Presence of many metastable states, separated by high barriers, is
a well-known problem in Monte Carlo simulations, involving
thermolization of large samples. The system easily gets trapped in
a metastable state and cannot efficiently explore the entire phase
space to find the ground state.\cite{mar01} The cooling process in
spin-glass experiments gives a similar result: the system is
trapped before the waiting time begins. However, Monte Carlo
studies of aging phenomena do not simulate the cooling protocol.
The simulations are started directly at the measurement
temperature from a random initial configuration, and results are
averaged over different initial conditions.
\cite{kis96,rit01,mar98,tak97,rit94} This may be the reason why
the subaging behavior is not normally observed in numerical
simulations. Moreover, the random initial configuration
corresponds to zero net magnetization. This means that Monte Carlo
studies are probably closer to the IRM, than to the TRM,
experiments.\cite{rit94} The difference between the two is not
very important, if memory of the initial state disappears quickly.
This, however, is not the case in real experiments.

The factorization of the TRM and IRM relaxation curves, Eq.~(15),
allows us to illustrate a possible origin of $\mu$ in a simple
way. The functions $f$ and $g$, Eqs.~(13) and (14), can be
approximated algebraically in the aging regime:
\begin{equation}
f \propto (t/ \, t_{w}^{\, \mu_{1}})^{-\lambda_{1}}~~;~~~~~~~~~ g
\propto (t/ \, t_{w}^{\, \mu_{2}})^{-\lambda_{2}}~~. \label{fit}
\end{equation}
It follows from Eq.~(15) that the relaxation exponent for the IRM
is $\lambda_{1}+\lambda_{2}$, while the exponent for the TRM is
$\lambda_{1}-\lambda_{2}$. The corresponding values of $\mu$
depend on the ratio $r=\lambda_{2}/\lambda_{1}$:
\begin{equation}
\mu_{IRM}=(\mu_{1}+\mu_{2} \, r)/ \, (1+r)~~; \label{mtr}
\end{equation}
\begin{equation}
\mu_{TRM}=(\mu_{1}-\mu_{2} \, r)/ \, (1-r)~~. \label{mir}
\end{equation}
The experimental values of $\mu_{1}$ and $\mu_{2}$ are exhibited
in Fig.~7. The exponents $\lambda_{1}$ are displayed in Fig.~9,
and the ratios $r=\lambda_{2}/\lambda_{1}$ -- in the inset of
Fig.~10.

\begin{figure}
\resizebox{\columnwidth}{!}{\includegraphics{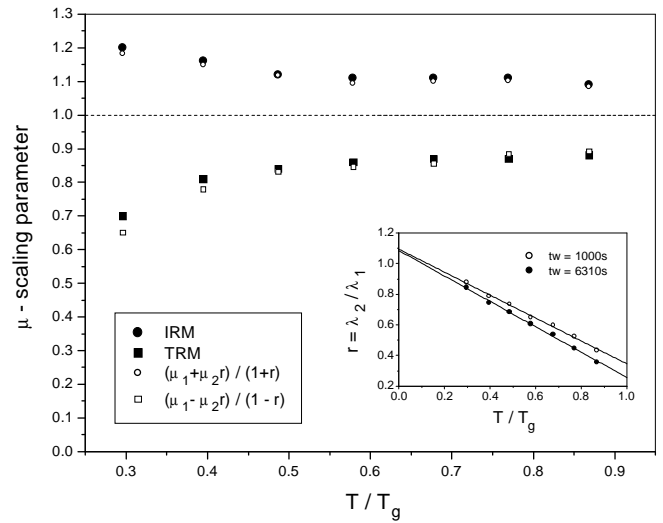}}
\caption{\label{F10} Illustration of a possible origin of $\mu$.
Parameters $\mu_{1}$ and $\lambda_{1}$ correspond to
$f=(\chi_{IRM} \, \chi_{TRM})^{1/2}$, while $\mu_{2}$ and
$\lambda_{2}$ correspond to $g=(\chi_{IRM}/\chi_{TRM})^{1/2}$.}
\end{figure}

The values of $\mu$ for the IRM and the TRM, determined from
Eqs.~(17) and (18), are shown in the main body of Fig.~10. They
compare well with the experimental values of $\mu$, obtained from
analysis of the actual relaxation curves. This comparison
demonstrates how the function $g$, which exhibits strong
superaging properties with $\mu_{2} \sim 1.2$, leads to $\mu \sim
0.9$ for the TRM, and $\mu \sim 1.1$ for the IRM. The results in
Fig.~10 were obtained using values of $r$ for $t_{w}=1000~s$. The
values of $r$ for $t_{w}=6310~s$ give similar numbers. For the
sake of simplicity, the observation time, $t$, is used in Eq.~(16)
instead of the $\mu$-dependent effective time, $t_{e}$, given by
Eq.~(9). The fact that Eqs.~(17) and (18) yield values of $\mu$
close to the their actual values suggests that this approach is
justified.

Appearance of the $\mu$-scaling can be explained in the following
way. Imagine that there is a function that exhibits good $t/t_{w}$
scaling. Let us also assume that effect of initial conditions is
described by another function, which does not scale well. This
second function will necessarily display superaging behavior,
approaching a constant as $t_{w}$ increases. Imagine also that
both these functions can be approximated by power laws of time,
which is often the case in spin-glass dynamics. If the first
function is fast, and the second function is slow, the resulting
decay will exhibit the $\mu$-scaling with $\mu$ not far from
unity. Of course, this is only one of possible scenarios.

According to Eqs.~(17) and (18), in order to achieve the best
scaling of both the TRM and the IRM (with $\mu$ close to unity),
one has to make the ratio $r=\lambda_{2}/\lambda_{1}$ as small as
possible. The inset of Fig.~10 suggests that $r$ decreases with
temperature and can be well approximated by a linear function of
$T/T_{g}$. It also diminishes with increasing $t_{w}$. Thus,
higher temperatures and longer waiting times should improve
$t/t_{w}$ scaling of measured relaxation curves.

Eq.~(18) also predicts that the quality of $\mu$-scaling of the
TRM curves should become worse as $r \rightarrow 1$ and $\mu_{1}
\rightarrow \mu_{2}$, because of the uncertainty 0/0. This happens
at low temperatures: the experimental values of $\mu$ for the TRM
drop sharply and their error bars increase considerably. The
physical reason for this is simple. The TRM relaxation at low
temperatures is completely dominated by the ZTRM, which is very
slow. The waiting time effects become almost indistinguishable,
and the $t/t_{w}$ scaling does not make much sense anymore. In
this case, the TRM is nearly constant, and both $f$ and $g$,
Eqs.~(13) and (14), are governed by the IRM. Therefore, the simple
average of the TRM and IRM does not give much better scaling at
low temperatures, as seen in Fig.~7.

These results can be interpreted in terms of an effective change
in barrier heights. Let us consider the logarithmic relaxation
rates: $S_{TRM}=-\partial \chi_{TRM}/\partial \ln(t)$ and
$S_{IRM}=-\partial \chi_{IRM}/\partial \ln(t)$. As usually, we
assume that both $\chi_{TRM}$ and $\chi_{IRM}$ are normalized by
one at $t=t_{w}$ before differentiation is performed. A ratio of
$S_{TRM}$ and $S_{IRM}$ can be regarded as a probability of
overcoming a barrier:
\begin{equation}
d \chi_{TRM}/ \, d \chi_{IRM} = \exp(-\Delta_{0} / k_{B}T)~~.
\label{del}
\end{equation}
This formula defines the effective barrier $\Delta_{0}(t,t_{w})$,
describing the difference in relaxation rates for two different
initial conditions.

Fig.~11 exhibits values of $\Delta_{0}/k_{B}T$ at $t=t_{w}$,
extracted from the experimental data according to Eq.~(19). One
can see that they grow steadily as temperature decreases. They are
also comparable with differences between the values of
$\ln(t_{w}^{e\!f\!f})$ for the TRM and the IRM. Values of
$\Delta_{0}(t_{w},t_{w})$ are displayed in the inset of Fig.~11.
They change only slightly over a wide range of temperatures, and
decrease with increasing $t_{w}$. The error bars for $\Delta_{0}$
are about 0.05.

\begin{figure}
\resizebox{\columnwidth}{!}{\includegraphics{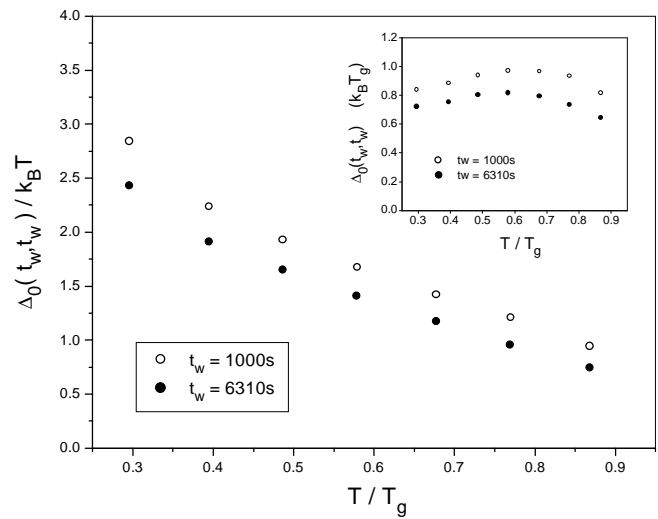}}
\caption{\label{F11} The effective barrier $\Delta_{0}$,
describing difference in relaxation rates for the TRM and the IRM,
at $t=t_{w}$. Note that it is almost independent of temperature,
as shown in the inset.}
\end{figure}

It has been shown experimentally that a change in a barrier
height, $\delta \Delta$, under a temperature change, $\delta T$,
is independent of temperature.\cite{ham92} The same appears to be
true for $\Delta_{0}$, which corresponds to an average change in
barrier heights due to the change in initial conditions. This fact
suggests that $\Delta_{0}$ is a result of cooling. Both the TRM
and the IRM are influenced by the cooling process, and it is
mainly this influence, that leads to systematic deviations from
full $t/t_{w}$ scaling in the aging regime. As the waiting time
increases, $\Delta_{0}$ diminishes, which means that memory of the
initial state is gradually washed away by the aging process.

It has been already mentioned that evolution of the system during
the waiting time $t_{w}$ can be characterized by the effective
barrier $\Delta(T,t_{w})=k_{B}T\ln(t_{w}/\tau_{0})$. The barriers
are raised by $\Delta_{TRM}$ for the TRM, and reduced by
$\Delta_{IRM}$ for the IRM. The average difference in barrier
heights for these experimental protocols is $\Delta_{0}=
\Delta_{TRM}+\Delta_{IRM}$. Using Eq.~(19), one can determine
$\Delta_{0}$ only. Fig.~8 suggests that $\Delta_{TRM}$ and
$\Delta_{IRM}$ may be different, but have similar properties.
Therefore, what is true for $\Delta_{0}$, is likely to be true for
$\Delta_{TRM}$ and $\Delta_{IRM}$ taken separately.

The characteristic barrier, $\Delta(T,t_{w})$, decreases with
decreasing temperature, because it corresponds to thermally
activated aging dynamics. The barrier $\Delta_{0}$ is almost
independent of temperature, presumably because it is related to
the cooling process. Therefore, departures from full $t/t_{w}$
scaling should become more pronounced as temperature is lowered.
Even if the cooling is fast, yielding small $\Delta_{0}$, it will
always be possible to find a sufficiently low measurement
temperature, $T$, so that $\Delta(T,t_{w}) \sim \Delta_{0}$, and
the $t/t_{w}$ scaling is strongly violated. According to this
argument, the sharp drop in $\mu$ for the TRM at low temperatures
is a natural consequence of cooling effects.\\

Some of the measurements, discussed above, have been repeated at
CEA Saclay. A SQUID susceptometer S600 by Cryogenics Ltd (UK) was
used for this purpose. The cooling process, which is a unique
characteristic of an apparatus, was somewhat different from the
one exhibited in Fig.~1. The experimental data, however, are
consistent with the results, reported in this paper.

\section{\label{summ}Conclusion}

There are many different approaches to analysis of $t/t_{w}$
scaling in spin glasses. In this paper, one more method is
proposed. It is based on the idea that the departures from
$t/t_{w}$ scaling, observed experimentally in the aging regime,
may reflect the influence of the cooling process. This means that
the thermal history of the spin-glass state cannot be neglected in
analysis of aging phenomena. The initial condition for aging is
not random in spin-glass experiments, and has a profound effect on
scaling properties of measured relaxation. It would be incorrect
to say that the cooling effects are spurious. They are as
important, as the aging phenomena. This is because they reflect
the hierarchical nature of spin-glass dynamics,\cite{bou02}
revealed by temperature-variation experiments.

The approach, used in this paper, is based on the detailed
comparison of time scaling properties of the TRM and the IRM. We
argue that, because of strong memory of the initial state, neither
of these functions represents the true $t_{w}$-dependent
relaxation. Even though they are characterized by the same waiting
time, their $t/t_{w}$ scaling properties are remarkably different.
We exploit this difference by taking an average of $\chi_{TRM}$
and $\chi_{IRM}$. This averaging over different initial conditions
leads to systematic improvement in $t/t_{w}$ scaling.
Factorization of measured relaxation functions allows us to
explain scaling properties of both the TRM and the IRM in a unique
way. Our results suggest that the observed deviations from
$t/t_{w}$ scaling in the aging regime are largely due to cooling
effects. The interrupted aging,\cite{bou94} which is another
possible reason for the lack of scaling, may be expected to play a
role at longer waiting times, especially in polycrystalline
samples.\\
\\

The conclusions, reached in this paper, are by no means final.
Their verification requires experiments in which different cooling
protocols in the immediate vicinity of the measurement temperature
can be implemented. Another option is a rapid field quench from
above the AT line. Monte Carlo studies, including simulations of
the cooling process in addition to aging effects, may also be very
helpful. The problem of $t/t_{w}$ scaling in spin-glass dynamics
deserves further experimental and theoretical investigation.\\

NOTE. After this paper had been completed, we received a preprint
by Berthier and Bouchaud,\cite{ber02} devoted to numerical studies
of the Edwards-Anderson model in 3 and 4 dimensions. One of their
conclusions is that ``a finite cooling rate effect...leads to an
apparent sub-aging behavior for the correlation function, instead
of the super-aging that holds for an infinitely fast quench''.
This is in agreement with the conclusion of the present paper that
$t/t_{w}$ scaling properties strongly depend on the initial
condition.\\
\\

\begin{acknowledgments}
We are most grateful to Professor~J.~A.~Mydosh for providing us
with the Cu:Mn single crystal sample, prepared in Kamerlingh Onnes
Laboratory (Leiden, The Netherlands). We thank Dr.~G.~G.~Kenning
for interesting discussions and valuable suggestions.\\
\end{acknowledgments}


\begin{thebibliography}{99}
\bibitem{comment} Note that a comprehensive
description of results from magnetic noise, relaxation, and
susceptibility measurements suggests that a quasiequilibrium
factor $(t^{-\alpha}+\delta)$, where $\delta$ is a constant,
should be used instead of $t^{-\alpha}$, as explained in Ref.~4
above.
\end{thebibliography}
\end{document}